\input{aipcheck}

\documentclass[
    ,final            
  ]
  {aipproc}

\layoutstyle{6x9}

\usepackage{journal_shortcuts}
\begin{document}

\title{Baryons in the outskirts of the X-ray brightest galaxy cluster}


\author{A. Simionescu}{
  address={KIPAC, Stanford University, 452 Lomita Mall, Stanford CA 94305, USA}
}

\author{S. W. Allen}{
  address={KIPAC, Stanford University, 452 Lomita Mall, Stanford CA 94305, USA}
}

\author{A. Mantz}{
  address={Kavli Institute for Cosmological Physics, University of Chicago, 5640 South Ellis Avenue, Chicago, IL 60637, USA}
  ,altaddress={NASA Goddard Space Flight Center, Greenbelt, MD 20771, USA} 
}
  
\author{N. Werner}{
  address={KIPAC, Stanford University, 452 Lomita Mall, Stanford CA 94305, USA}
}

\author{Y. Takei}{
  address={Institute of Space and Astronautical Science (ISAS), JAXA, 3-1-1 Yoshinodai, Sagamihara, Kanagawa 229-8510, Japan}
}

\begin{abstract}
Studies of the diffuse X-ray emitting gas in
   galaxy clusters have provided powerful constraints on
   cosmological parameters and insights into plasma astrophysics.
   However, measurements of the faint cluster outskirts have become
   possible only over the last few years. 
  Here, we present results from Suzaku observations of the Perseus Cluster, which provide our best measurements of the thermodynamic properties of the ICM at large
   radii to date. In particular, we focus on the details of the data analysis procedure and discuss the evidence for a clumpy distribution
   of the gas in the outskirts, which is important for understanding
   the physics of the ongoing growth of clusters from the
   surrounding cosmic web.
   \end{abstract}

\maketitle


\section{Motivation}

In the hierarchical picture of large-scale structure formation in the Universe, clusters of galaxies, which are the most massive objects, are also the latest to form. Galaxy clusters are enormous knots in the cosmic web, located at the intersections of large-scale structure filaments, from which they are still accreting matter at the present time. Detailed observations of cluster outskirts therefore allow us not only to determine the cluster properties accurately but also to witness and understand large-scale structure formation as it happens. 

The positions of clusters at the high end of the mass spectrum makes them particularly useful and sensitive cosmological probes, provided that their total mass can be measured accurately \citep[for a review, see][]{Allen11}. Hydrostatic mass measurements extending to large radii, used in combination with other thermodynamical X-ray observables and measurements of weak-lensing, galaxy velocity dispersion, and the Sunyaev-Zel'dovich (SZ) effect, provide critical constraints on structure formation models and a robust comparison point for simulations.

Observations of cluster outskirts can be used to search for the boundary between virialized and infalling material, beyond which the hydrostatic equilibrium assumption breaks down. 
At larger radii, we can therefore expect to detect inhomogeneities in the ICM, such as infalling matter clumps and signatures of accretion shocks, which could provide a direct view of the physical processes by which clusters grow.
Given even a limited resolution in the azimuthal direction we can furthermore ask whether the signatures of gas accretion are ubiquitous or occur preferentially along the major axis or in the direction of known large-scale structure filaments in supercluster environments.

\section{The role of the Suzaku satellite}

Until recently, only a limited amount of information regarding the X-ray surface brightness of cluster outskirts was available, primarily based on {\it ROSAT} observations \citep{Vikhlinin99,Neumann05}.
The low surface brightness signal and relatively high instrumental noise of the detectors onboard {\it Chandra} and {\it XMM-Newton} have limited robust measurements of the temperature of the hot, diffuse intracluster medium (ICM) to radii less than approximately half of the virial radius \citep[e.g.][]{vikhlinin2006}. Implicitly, measurements of the total cluster mass, determined under the assumption of hydrostatic equilibrium, could not be extended outside half of the virial radius except with the use of parametrized extrapolations. 

The lower and more stable particle background of {\it Suzaku} has enabled a breakthrough in determining the ICM properties out to the virial radius in several bright, relatively nearby clusters \citep{George09,Reiprich09,Bautz09,Hoshino10,Kawaharada10,Akamatsu11}. In all cases, the {\it Suzaku} results point towards temperature profiles which decline smoothly from a few hundred kpc out to large radii, which is qualitatively in agreement with predictions from numerical simulations \citep[e.g.][]{Evrard96}. For several clusters, it was found that, at large radii, the entropy profile flattens compared to the power-law increase with radius observed in the central regions \citep{George09,Akamatsu11}, possibly pointing towards the onset of non-hydrostatic equilibrium effects and convective instability around and beyond the virial radius. 

\section{The outskirts of the Perseus Cluster: an ideal target for Suzaku}

The Perseus Cluster of galaxies (A426) is the brightest, extended X-ray source. It is both closer (at z=0.0183) than any of the clusters for which {\it Suzaku} has been previously used to measure the ICM properties at large radii, and it has a significantly higher X-ray flux. The large angular size of the cluster mitigates the need to precisely model the effects of {\it Suzaku}'s complex point spread function (PSF), making the Perseus cluster an ideal target to study cluster outskirts. Furthermore, the large angular size permits the use of larger extraction regions for the same fractional interval of the virial radius, which translates into collecting signal faster and reducing the expected root mean squared variations of the background point source flux. A wealth of information is already available about the central parts of the Perseus cluster, from the Uhuru and Einstein satellites up to a recent 1.4 megasecond Chandra image \citep{Fabian11}. This information, in combination with the unprecedented {\it Suzaku} measurements of the temperature out to the virial radius which will be presented in the following sections, provides the most detailed view of a galaxy cluster ever achieved. 

\subsection{The pilot project on the outskirts of the Perseus Cluster}

Between 2009 July 29 and 2009 August 22, Suzaku observed a mosaic of 14 fields placed along two arms towards the east and northwest from NGC1275, the central galaxy of the Perseus cluster. NGC1275 is monitored regularly as a calibration source, therefore in addition we used one central pointing performed on 2009 August 26, which is as close in time as possible to our mosaic fields and was available from the public archive. We focus on the data obtained with the three available X-ray imaging spectrometer (XIS) cameras. {\it The main results on this study are described in \cite{SimionescuSci}. The purpose of the current manuscript is to complement the main journal article and supply additional information regarding the details of the data analysis performed}.

\begin{figure}\label{sb}
  \includegraphics[height=.65\textwidth]{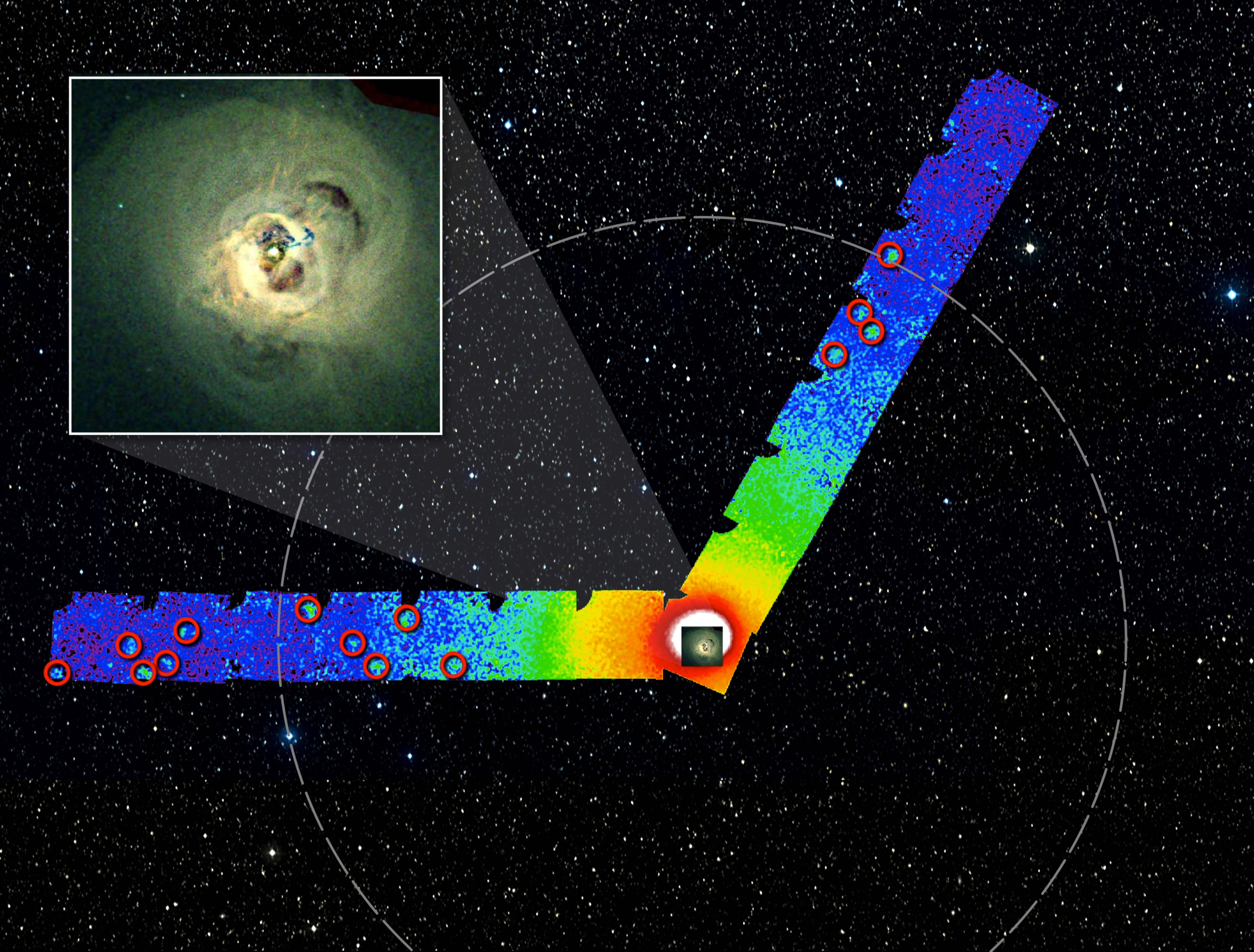}
  \caption{The X-ray surface brightness map of the NW and E arms of the Perseus Cluster, superimposed to scale on an optical SDSS image. The dashed circle marks $r{200}$. The inset shows a deep Chandra image of the cluster's bright central region.}
\end{figure}

The data were reduced using the tools available in the HEAsoft package (version 6.9) to create a set of cleaned event lists with hot or flickering pixels removed. All the standard recommended screening criteria were applied (see e.g. \url{http://heasarc.nasa.gov/docs/suzaku/processing/criteria_xis.html}) and we selected times with the geomagnetic cut-off rigidity COR $>$ 6 GV. We also verified using the method described in \cite{Yoshino09} that the geometry of Earth's magnetic field was not favorable for generating strong solar wind charge exchange (SWCX) signatures during our observations.  

The X-ray surface brightness image, extracted in the 0.7--7 keV energy band and corrected for vignetting and instrumental background, is shown in two false-color strips in Fig. \ref{sb}. We visually identified bright point sources marked by red circles in the figure. 

\subsubsection{Background analysis}\label{sect:bkg}
The non X-ray background (NXB) spectrum was constructed from the dark Earth database using the standard method in which the COR distributions of the on-source and the background data are matched. The cosmic X-ray background (CXB) is typically modeled with three components: a thermal component with kT$ \sim 0.09$ keV to account for the local hot bubble (LHB) emission, a second thermal component with kT$ \sim 0.2$ keV for the Galactic halo (GH), and a power-law with $\Gamma=1.41$ to account for the integrated emission of unresolved point sources \citep[e.g.][]{kuntz2000,snowden2008}. Additionally, we include a thermal component with kT$\sim 0.6-0.8$ keV (hot foreground, HF), which is sometimes required to describe the Galactic foreground, particularly at lower Galactic latitudes. 

Data from the Suzaku mosaic beyond the estimated virial radius of Perseus (the outer two pointings along each arm), as well as from the ROSAT All Sky Survey (RASS) can be used to constrain the parameters of the background model and estimate their spatial variations. 
From RASS, we used 5 circles with a radius of 1 degree located $3^\circ \: 10^\prime$ to the N, NE, E, SE, and S of the cluster center, avoiding contamination from Algol and AWM7 on the western side.
Solar abundances were assumed for all thermal components \cite{Feldman92}. Table \ref{cxb} shows an excellent agreement between ROSAT and Suzaku for all the model parameters. For our final CXB model, we chose the best-fit parameters of the GH, HF, and power-law based on the Suzaku data. The LHB parameters, which cannot be constrained with Suzaku, were fixed based on the ROSAT data.

To determine the spatial variations of the Galactic foreground, we fitted spectra from the RASS circles leaving the normalizations of the GH, HF and LHB components to vary independently for all 5 spectra. The temperatures of all thermal components were fixed to our choice of the final CXB model parameters described above. The LHB contributes very little to the emission above 0.5 keV, thus we do not calculate the effect of systematic uncertainties on the LHB parameters. To estimate the systematic uncertainties on the CXB power-law, we fitted 8 Suzaku background spectra (each of the outer two pointings along the two arms was divided in half) individually in the 2--7 keV band and computed the variance of the 8 values of the normalization. The area of each of the 8 regions used was approximately 160 sq arcmin, thus for smaller spectral extraction regions of area $A$ the variance is expected to increase by $\sqrt{160\: {\rm arcmin^2}/A}$. 

\begin{table}
\caption{Cosmic X-ray Background parameters. The temperatures are given in keV, all normalizations represent the standard XSPEC value normalized to $10^{-3}$ per $20^2\pi$ sq arcmin area. Errors are given at the $\Delta$C=1 level.}
\begin{tabular}{lcccc}
\hline
\hline
& Parameter	& Suzaku  & ROSAT & Systematic uncertainty\\
\hline
LHB  & kT         &  0.091 (fix)  & $0.091\pm0.003$ & -- \\
          & norm    &  1.70 (fix) & $1.70\pm0.07$ & -- \\
GH    & kT         & $0.18\pm0.02$ & $0.17^{+0.04}_{-0.07}$ & -- \\
          & norm    & $1.59^{+0.43}_{-0.28}$ & $2.06^{+0.29}_{-0.38}$ & 1.43 (S) -- 2.91 (NE) \\
HF   & kT         & $0.615\pm0.036$    & 0.60 (fix) & -- \\
         & norm    & $0.52\pm0.07$         & $0.50\pm0.07$ & 0.31(S) -- 0.60 (SE) \\
Power-law & $\Gamma$ &  1.41 (fix) & 1.41 (fix) & -- \\
      & norm    & $1.096\pm0.016$& 1.138 (fix) & 9.4\% $\times \sqrt{160\: {\rm arcmin^2}/A}$ \\
\hline
\end{tabular}
\label{cxb}
\end{table}

\subsubsection{Stray-light rejection for the spectral analysis}

For each annulus, we calculated the difference between the observed spectrum and the assumed CXB model presented above. This yields a background-subtracted cluster signal.
We then divided by this signal the corresponding stray light spectrum extracted from the event files generated by our ray-tracing simulations. We included in the spectral fitting only the parts of each observed spectrum between 0.7--7 keV where the contamination was below 20\%. Because the stray light spectrum peaks between 1--1.2 keV, even in regions with high stray-light contamination the spectral contribution of the stray light above 1.5 keV becomes typically less than 10\%, and without exception less than 20\%. 

\subsubsection{Results}

\begin{figure}\label{sys}
  \includegraphics[height=.5\textwidth]{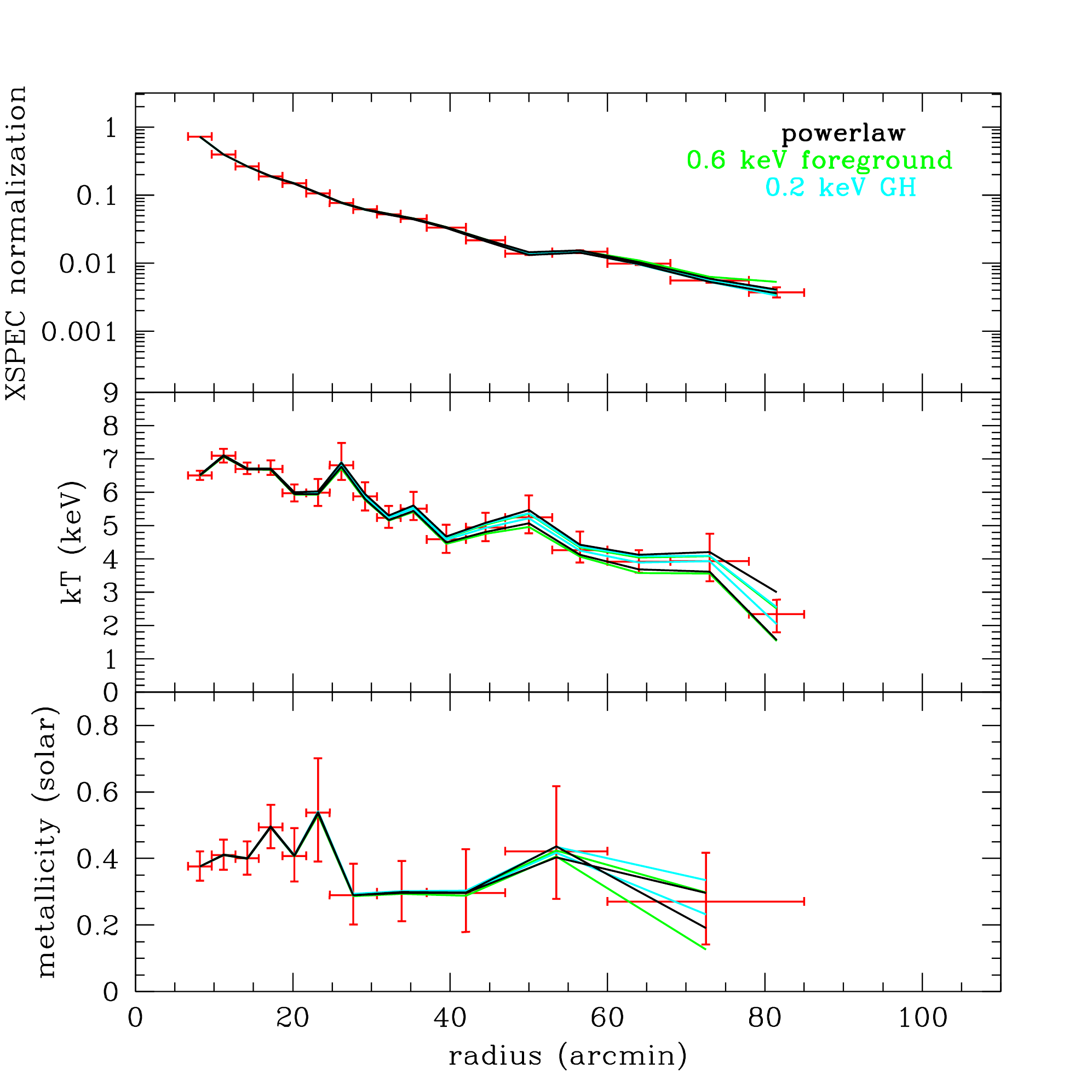}
    \includegraphics[height=.5\textwidth]{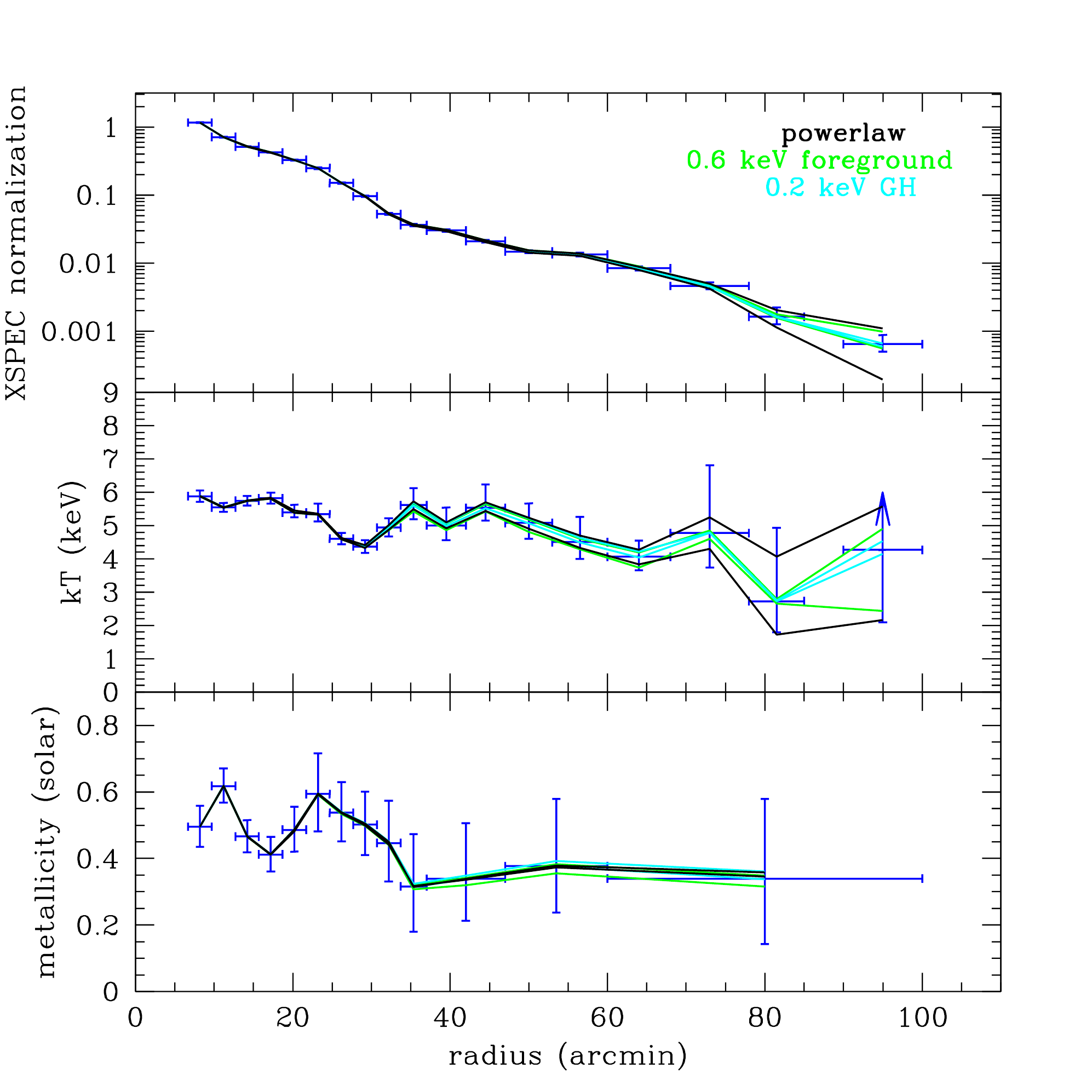}
  \caption{Effects of systematic uncertainties in the CXB parameters on the cluster normalization and projected temperature and metallicity profiles.}
\end{figure}

Combining our results with those obtained from an ultra-deep Chandra observation of the cluster center\cite{sanders2007}, we find that the Suzaku and Chandra radial profiles show excellent agreement where they intersect, and together measure the temperature and metallicity structure of the intra-cluster gas with high precision and previously unattained spatial resolution out to the virial radius ($r_{200}$) \cite{SimionescuSci}. In the narrow interval spanning 0.95-1.05$r_{200}$, the temperature is approximately a third of the peak temperature. The systematic errors introduced by the CXB modeling are typically smaller than the statistical $1\sigma$ errors, as demonstrated in Fig. \ref{sys}. 

\begin{figure}\label{ne}
  \includegraphics[width=.45\textwidth]{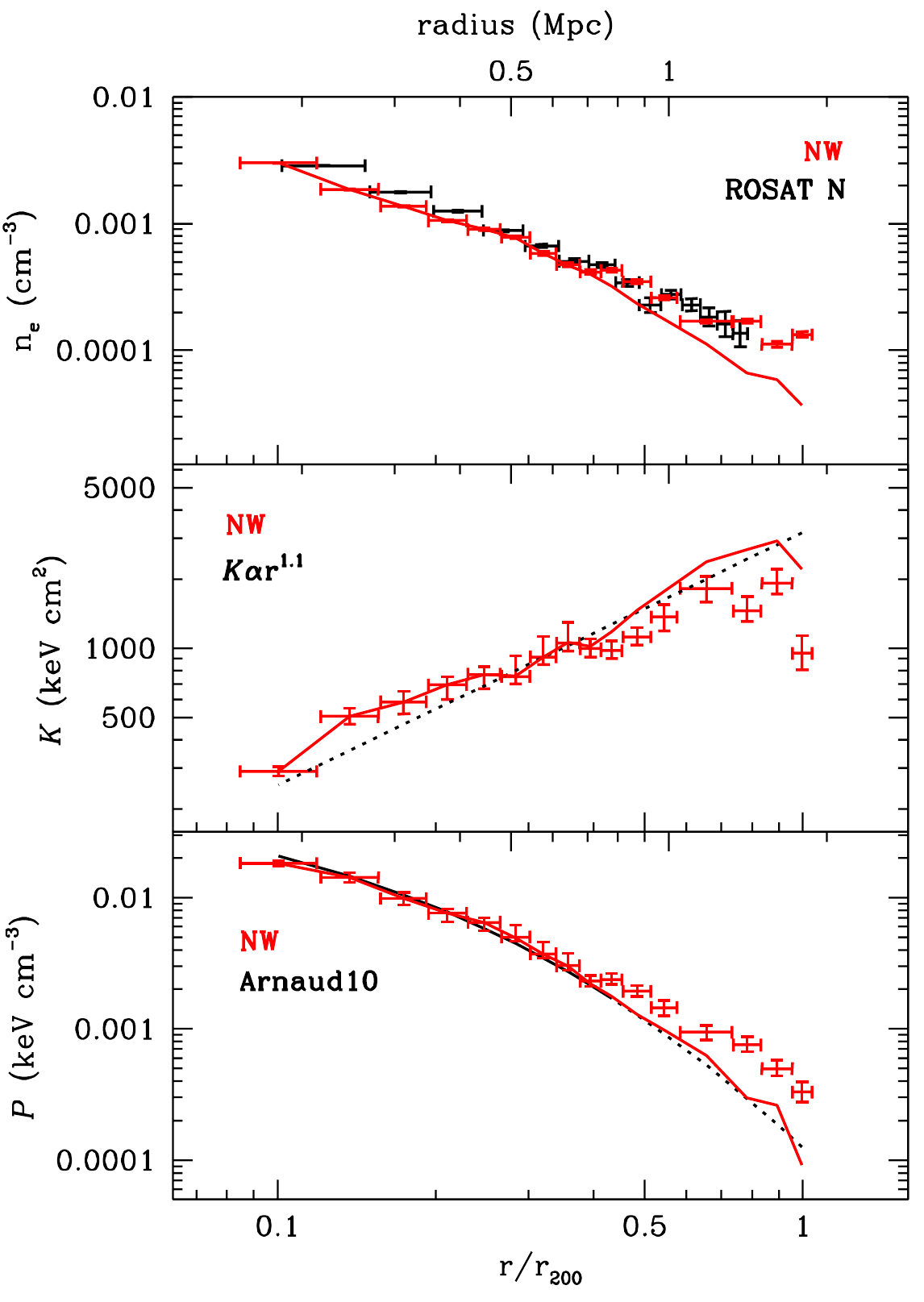}
  \includegraphics[width=.55\textwidth]{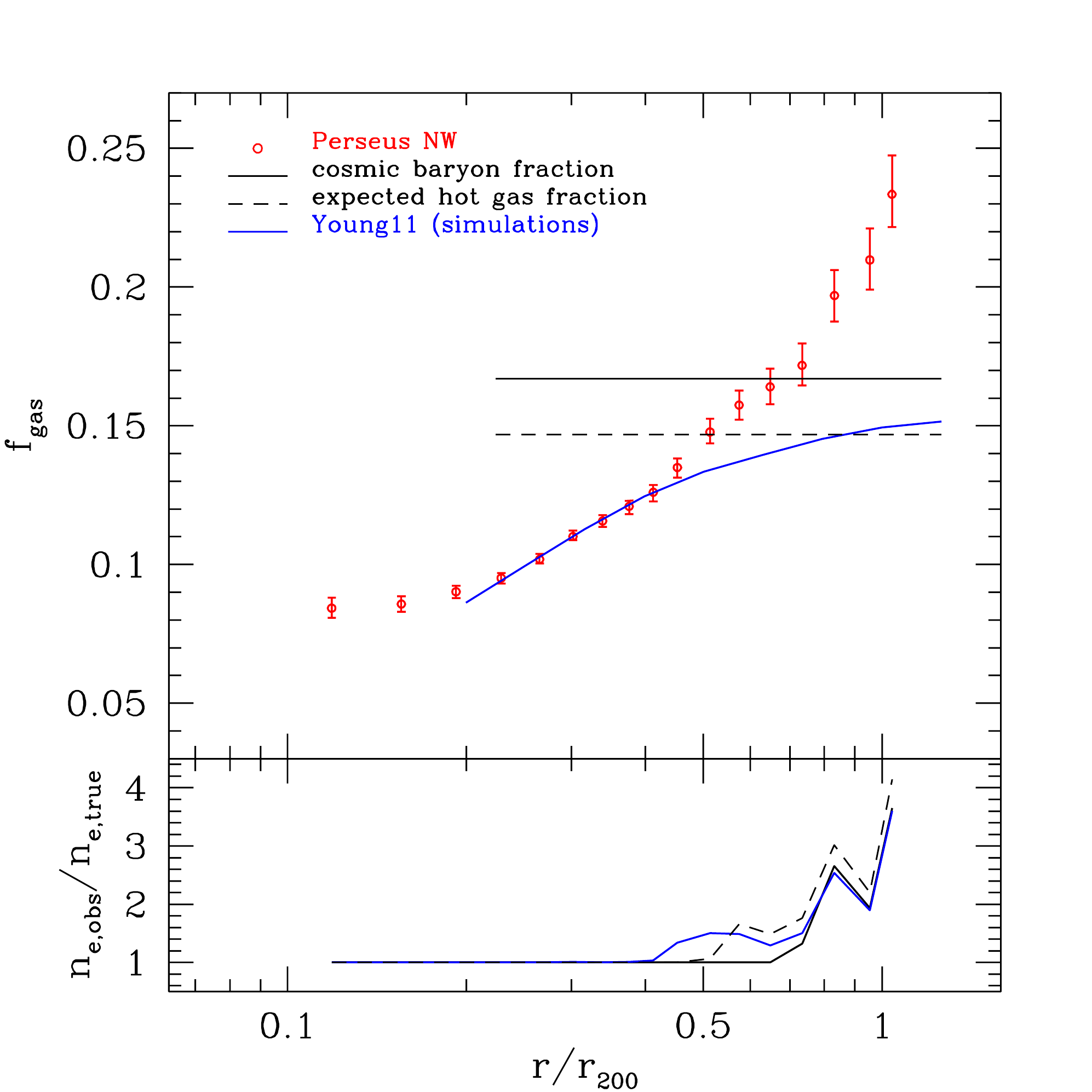}
  \caption{{\it Left:} The deprojected electron density ($n_e$), entropy ($K$), and pressure ($P$) profiles measured with Suzaku along the NW arm \cite{SimionescuSci} are shown with red data points. The top panel shows the agreement with the $n_e$ measured with ROSAT in the N quadrant \cite{Ettori98}. The red line shows the same profiles corrected for clumping. The expected entropy profile from simulations of gravitational collapse\cite{voit2005MNRAS} is a power-law with index $\beta\sim1.1$, over-plotted as a black dotted line in the entropy panel. The average profile of a sample of clusters previously studied with the XMM-Newton satellite within $\sim$0.5$r_{200}$\cite{arnaud2010} is shown with a solid black curve in the pressure panel; its extrapolation to $r_{200}$ is shown with a dotted black line.
 {\it Right:} The integrated, enclosed gas mass fraction profile for the NW arm. The cosmic baryon fraction from WMAP7\cite{komatsu2010} is indicated by the horizontal solid black line; accounting for 12\% of the baryons being in stars gives the expected fraction of baryons in the hot gas phase, shown as a dashed black line. Predictions from numerical simulations\cite{young2010} are shown in blue. The bottom panel shows the inferred clumping factor.
  }
\end{figure}

We are also able to obtain exquisite deprojected radial profiles of the electron density, entropy, and pressure. Assuming that the total mass distribution follows an NFW profile, as suggested by numerical simulations \cite{navarro1997}, we used the data from the dynamically relaxed northwestern arm to determine the best-fit total mass profile and obtain an unprecedently accurate gas mass fraction profile, which extends out to the virial radius of the Perseus Cluster (shown in the right panel of Fig. \ref{ne}).

This profile reveals no evidence for the puzzling deficit of baryons at $r \ge 0.5 r_{200}$ inferred from some previous studies of other systems, using lower quality data and/or extrapolated models
\citep[e.g.][]{afshordi2007,vikhlinin2006}. At $r_{500}$, the cluster has the expected (approximately Universal) baryon fraction. Within $r < 0.5 r_{200}$, then, the physics of the X-ray emitting gas appears relatively simple and X-ray measurements can be used robustly for cosmological work. At larger radii, the apparent gas mass fraction exceeds the cosmic mean baryon fraction measured from the CMB \cite{komatsu2010}. This excess also correlates with an apparent flattening of the entropy profile. These results are robust against all known systematic uncertainties. The most plausible explanation for these properties is that the gas at large radii is clumpy\footnote{if the density is non-uniform, then the average of the square of the electron density, which determines the X-ray bremsstrahlung emission, can be larger than the square of the average electron density, $\left<n_e^2\right> > \left<n_e\right>^2$. Determining the electron density from X-ray observations as $\sqrt{\left<n_e^2\right>}$ will overestimate the true $n_e$.}. The deprojected radial profile of the electron density shows excellent agreement with previous ROSAT measurements extending out to $\sim1.4$ Mpc\cite{Ettori98}, therefore we can exclude any residual stray light contamination in the Suzaku data as a possible reason for the boosted electron density at large radii.

Outside the central region, and inside the radius where clumping becomes important, the measured $f_{gas}$ profile shows good agreement with recent numerical simulations\citep{young2010}, where a semi-analytic model was used to calculate the energy transferred to the intra-cluster gas by supernovae and AGN during the galaxy formation process. Extrapolating this model into the outskirts where clumping is important, we used its predictions together with the measured $f_{gas}(r)$ to determine by how much the electron density must be overestimated to produce the difference between the data and the model. Correcting the electron density using this factor, we find that the entropy profile becomes consistent with the power-law profile expected from models of gravitational structure formation \citep{voit2005MNRAS}. Additionally, the pressure becomes consistent with that expected by extrapolating the average profile of a sample of clusters previously studied with XMM-Newton\citep{arnaud2010}; this ``universal'' pressure profile is also motivated by numerical simulations \cite{Nagai07}. Correcting for the bias introduced by gas clumping is the only straightforward way to bring {\it both} the entropy {\it and} pressure profiles in agreement with the expected models.

Numerical simulations predict gas clumping in the cluster outskirts \cite{Roncarelli06,Nagai11}, but the amount of clumping depends on many physical processes in the ICM which are currently uncertain (e.g. viscosity, conduction, star formation feedback, magnetic fields), and on the numerical schemes used \citep{Vazza11}. To explore the implications of gas clumping for cluster astrophysics, we must combine efforts in both observation and theory. 

\section{The Perseus Cluster Key Project and future outlook}

Numerical simulations predict large variations in the gas clumping factor as a function of azimuth and dynamical state of the ICM. It is therefore crucial to have measurements of similar precision to our  pilot project along other directions in Perseus, as well as in other systems. The Perseus Cluster Key Project will extend the existing mosaic by 6 additional arms (2 in AO-5 and 4 in AO-6), providing a uniform sampling in azimuth. In addition, Key and Large projects covering the Coma Cluster (a very massive, but unrelaxed system) and A2199 (a less massive, relaxed cluster) will provide crucial information about the variation of the thermodynamic and $f_{gas}$ profiles and of the clumping factor in systems of different masses and dynamical states. High-resolution imaging with Chandra will further reveal the shapes and size distribution of the brightest gas clumps.

\bibliographystyle{aipproc}
\bibliography{bibliography,clustersnewest,clustersvirial}

\begin{thebibliography}{29}
\expandafter\ifx\csname natexlab\endcsname\relax\def\natexlab#1{#1}\fi
\providecommand{\enquote}[1]{``#1''}
\expandafter\ifx\csname url\endcsname\relax
  \def\url#1{\texttt{#1}}\fi
\expandafter\ifx\csname urlprefix\endcsname\relax\def\urlprefix{URL }\fi
\providecommand{\eprint}[2][]{\url{#2}}

\bibitem[{Allen} et~al.(2011)]{Allen11}
S.~W. {Allen}, A.~E. {Evrard}, and A.~B. {Mantz}, \emph{\araa} \textbf{49},
  409--470 (2011), \eprint{1103.4829}.

\bibitem[{Vikhlinin} et~al.(1999)]{Vikhlinin99}
A.~{Vikhlinin}, W.~{Forman}, and C.~{Jones}, \emph{\apj} \textbf{525}, 47--57
  (1999), \eprint{arXiv:astro-ph/9905200}.

\bibitem[{Neumann}(2005)]{Neumann05}
D.~M. {Neumann}, \emph{\aap} \textbf{439}, 465--477 (2005),
  \eprint{arXiv:astro-ph/0505049}.

\bibitem[{Vikhlinin} et~al.(2006)]{vikhlinin2006}
A.~{Vikhlinin}, A.~{Kravtsov}, W.~{Forman}, C.~{Jones}, M.~{Markevitch}, S.~S.
  {Murray}, and L.~{Van Speybroeck}, \emph{\apj} \textbf{640}, 691--709 (2006),
  \eprint{arXiv:astro-ph/0507092}.

\bibitem[{George} et~al.(2009)]{George09}
M.~R. {George}, A.~C. {Fabian}, J.~S. {Sanders}, A.~J. {Young}, and H.~R.
  {Russell}, \emph{\mnras} \textbf{395}, 657--666 (2009), \eprint{0807.1130}.

\bibitem[{Reiprich} et~al.(2009)]{Reiprich09}
T.~H. {Reiprich}, D.~S. {Hudson}, Y.~{Zhang}, K.~{Sato}, Y.~{Ishisaki},
  A.~{Hoshino}, T.~{Ohashi}, N.~{Ota}, and Y.~{Fujita}, \emph{\aap}
  \textbf{501}, 899--905 (2009), \eprint{0806.2920}.

\bibitem[{Bautz} et~al.(2009)]{Bautz09}
M.~W. {Bautz}, E.~D. {Miller}, J.~S. {Sanders}, K.~A. {Arnaud}, R.~F.
  {Mushotzky}, F.~S. {Porter}, K.~{Hayashida}, J.~P. {Henry}, J.~P. {Hughes},
  M.~{Kawaharada}, K.~{Makashima}, M.~{Sato}, and T.~{Tamura}, \emph{\pasj}
  \textbf{61}, 1117-- (2009), \eprint{0906.3515}.

\bibitem[{Hoshino} et~al.(2010)]{Hoshino10}
A.~{Hoshino}, J.~{Patrick Henry}, K.~{Sato}, H.~{Akamatsu}, W.~{Yokota},
  S.~{Sasaki}, Y.~{Ishisaki}, T.~{Ohashi}, M.~{Bautz}, Y.~{Fukazawa},
  N.~{Kawano}, A.~{Furuzawa}, K.~{Hayashida}, N.~{Tawa}, J.~P. {Hughes},
  M.~{Kokubun}, and T.~{Tamura}, \emph{\pasj} \textbf{62}, 371-- (2010),
  \eprint{1001.5133}.

\bibitem[{Kawaharada} et~al.(2010)]{Kawaharada10}
M.~{Kawaharada}, N.~{Okabe}, K.~{Umetsu}, M.~{Takizawa}, K.~{Matsushita},
  Y.~{Fukazawa}, T.~{Hamana}, S.~{Miyazaki}, K.~{Nakazawa}, and T.~{Ohashi},
  \emph{\apj} \textbf{714}, 423--441 (2010), \eprint{1002.4811}.

\bibitem[{Akamatsu} et~al.(2011)]{Akamatsu11}
H.~{Akamatsu}, A.~{Hoshino}, Y.~{Ishisaki}, T.~{Ohashi}, K.~{Sato}, Y.~{Takei},
  and N.~{Ota}, \emph{ArXiv e-prints}  (2011), \eprint{1106.5653}.

\bibitem[{Evrard} et~al.(1996)]{Evrard96}
A.~E. {Evrard}, C.~A. {Metzler}, and J.~F. {Navarro}, \emph{\apj} \textbf{469},
  494--+ (1996), \eprint{arXiv:astro-ph/9510058}.

\bibitem[{Fabian} et~al.(2011)]{Fabian11}
A.~C. {Fabian}, J.~S. {Sanders}, S.~W. {Allen}, R.~E.~A. {Canning},
  E.~{Churazov}, C.~S. {Crawford}, W.~{Forman}, J.~{GaBany},
  J.~{Hlavacek-Larrondo}, R.~M. {Johnstone}, H.~R. {Russell}, C.~S. {Reynolds},
  P.~{Salome}, G.~B. {Taylor}, and A.~J. {Young}, \emph{ArXiv e-prints}
  (2011), \eprint{1105.5025}.

\bibitem[{Simionescu} et~al.(2011)]{SimionescuSci}
A.~{Simionescu}, S.~W. {Allen}, A.~{Mantz}, N.~{Werner}, Y.~{Takei}, R.~G.
  {Morris}, A.~C. {Fabian}, J.~S. {Sanders}, P.~E.~J. {Nulsen}, M.~R. {George},
  and G.~B. {Taylor}, \emph{Science} \textbf{331}, 1576-- (2011),
  \eprint{1102.2429}.

\bibitem[{Yoshino} et~al.(2009)]{Yoshino09}
T.~{Yoshino}, K.~{Mitsuda}, N.~Y. {Yamasaki}, Y.~{Takei}, T.~{Hagihara},
  K.~{Masui}, M.~{Bauer}, D.~{McCammon}, R.~{Fujimoto}, Q.~D. {Wang}, and
  Y.~{Yao}, \emph{\pasj} \textbf{61}, 805-- (2009), \eprint{0903.2981}.

\bibitem[{Kuntz} and {Snowden}(2000)]{kuntz2000}
K.~D. {Kuntz}, and S.~L. {Snowden}, \emph{\apj} \textbf{543}, 195--215 (2000).

\bibitem[{Snowden} et~al.(2008)]{snowden2008}
S.~L. {Snowden}, R.~F. {Mushotzky}, K.~D. {Kuntz}, and D.~S. {Davis},
  \emph{\aap} \textbf{478}, 615--658 (2008), \eprint{arXiv:0710.2241}.

\bibitem[{Feldman}(1992)]{Feldman92}
U.~{Feldman}, \emph{\physscr} \textbf{46}, 202--220 (1992).

\bibitem[{Sanders} and {Fabian}(2007)]{sanders2007}
J.~S. {Sanders}, and A.~C. {Fabian}, \emph{\mnras} \textbf{381}, 1381--1399
  (2007), \eprint{0705.2712}.

\bibitem[{Navarro} et~al.(1997)]{navarro1997}
J.~F. {Navarro}, C.~S. {Frenk}, and S.~D.~M. {White}, \emph{\apj} \textbf{490},
  493--+ (1997), \eprint{arXiv:astro-ph/9611107}.

\bibitem[{Afshordi} et~al.(2007)]{afshordi2007}
N.~{Afshordi}, Y.-T. {Lin}, D.~{Nagai}, and A.~J.~R. {Sanderson}, \emph{\mnras}
  \textbf{378}, 293--300 (2007), \eprint{arXiv:astro-ph/0612700}.

\bibitem[{Komatsu} et~al.(2011)]{komatsu2010}
E.~{Komatsu}, K.~M. {Smith}, J.~{Dunkley}, C.~L. {Bennett}, B.~{Gold},
  G.~{Hinshaw}, N.~{Jarosik}, D.~{Larson}, M.~R. {Nolta}, L.~{Page}, D.~N.
  {Spergel}, M.~{Halpern}, R.~S. {Hill}, A.~{Kogut}, M.~{Limon}, S.~S. {Meyer},
  N.~{Odegard}, G.~S. {Tucker}, J.~L. {Weiland}, E.~{Wollack}, and E.~L.
  {Wright}, \emph{\apjs} \textbf{192}, 18--+ (2011), \eprint{1001.4538}.

\bibitem[{Ettori} et~al.(1998)]{Ettori98}
S.~{Ettori}, A.~C. {Fabian}, and D.~A. {White}, \emph{\mnras} \textbf{300},
  837--856 (1998), \eprint{arXiv:astro-ph/9806375}.

\bibitem[{Voit} et~al.(2005)]{voit2005MNRAS}
G.~M. {Voit}, S.~T. {Kay}, and G.~L. {Bryan}, \emph{\mnras} \textbf{364},
  909--916 (2005), \eprint{arXiv:astro-ph/0511252}.

\bibitem[{Arnaud} et~al.(2010)]{arnaud2010}
M.~{Arnaud}, G.~W. {Pratt}, R.~{Piffaretti}, H.~{B{\"o}hringer}, J.~H.
  {Croston}, and E.~{Pointecouteau}, \emph{\aap} \textbf{517}, A92 (2010),
  \eprint{0910.1234}.

\bibitem[{Young} et~al.(2011)]{young2010}
O.~E. {Young}, P.~A. {Thomas}, C.~J. {Short}, and F.~{Pearce}, \emph{\mnras}
  \textbf{413}, 691--704 (2011), \eprint{1007.0887}.

\bibitem[{Nagai} et~al.(2007)]{Nagai07}
D.~{Nagai}, A.~V. {Kravtsov}, and A.~{Vikhlinin}, \emph{\apj} \textbf{668},
  1--14 (2007), \eprint{arXiv:astro-ph/0703661}.

\bibitem[{Roncarelli} et~al.(2006)]{Roncarelli06}
M.~{Roncarelli}, S.~{Ettori}, K.~{Dolag}, L.~{Moscardini}, S.~{Borgani}, and
  G.~{Murante}, \emph{\mnras} \textbf{373}, 1339--1350 (2006),
  \eprint{arXiv:astro-ph/0609824}.

\bibitem[{Nagai} and {Lau}(2011)]{Nagai11}
D.~{Nagai}, and E.~T. {Lau}, \emph{\apjl} \textbf{731}, L10+ (2011),
  \eprint{1103.0280}.

\bibitem[{Vazza} et~al.(2011)]{Vazza11}
F.~{Vazza}, K.~{Dolag}, D.~{Ryu}, G.~{Brunetti}, C.~{Gheller}, H.~{Kang}, and
  C.~{Pfrommer}, \emph{ArXiv e-prints}  (2011), \eprint{1106.2159}.

\end{thebibliography}

\end{document}